\documentclass[]{llncs}

\pdfoutput=1
\usepackage[utf8]{inputenc}
\usepackage[T1]{fontenc}

\usepackage[usenames,dvipsnames]{xcolor}
\usepackage{amssymb,amsmath,amsfonts}
\usepackage{tikz,pgffor}
\usetikzlibrary{arrows}
\usetikzlibrary{shapes}
\usetikzlibrary{calc}
\usetikzlibrary{automata}
\usepackage{txfonts}
\usepackage{wrapfig}
\usepackage{cutwin}
\usepackage{stackrel}
\usepackage{enumitem}
\usepackage{eurosym}
\usepackage{listings}
\usepackage{multirow}

\usepackage{multirow}

\usepackage{hyperref}

\usepackage{microtype}

\newcommand{\theoremlike}[2]{\par\medskip\penalty-250\refstepcounter{theorem}{{\bfseries\noindent#2
\ref{#1}.}}}

\newcommand{\thmhelperpre}[2]{\theoremlike{#1}{#2}}
\newcommand{\thmhelperpost}{\par\medskip}

 \advance\textwidth0.7mm
 \newcommand{\zmensovatkoVertMezer}{0pt}

\newcommand{\Nset}{\mathbb{N}}
\newcommand{\Nseto}{\Nset_0}

\newcommand{\Rsetp}{\mathbb{R}_{>0}}
\newcommand{\Rsetpo}{\mathbb{R}_{\ge 0}}

\newcommand{\eps}{\varepsilon}
\newcommand{\dist}{\mathcal{D}}

\newcommand{\expred}{E}

\newcommand{\fdC}{C}

\newcommand{\states}{S}
\newcommand{\sta}{s}

\newcommand{\goalStates}{T}

\newcommand{\initstate}{\sta_{in}}
\newcommand{\fd}{fd }

\newcommand{\sleep}{\mathit{sleep}}
\newcommand{\busy}{\mathit{busy}}
\newcommand{\idle}{\mathit{idle}}

\newcommand{\paragraphskip}{\medskip}

\newcommand{\delays}{d}

\newcommand{\fDSet}{F}
\newcommand{\rates}{Q}
\newcommand{\occur}{N}
\newcommand{\sched}{A}
\newcommand{\expEvent}{\mathcal{E}}

\newcommand{\rateRew}{\mathcal{R}}
\newcommand{\impRew}{\mathcal{I}}

\newcommand{\costFdC}{\mathit{Cost}_{\fdC,\goalStates,\rateRew,\impRew}}

\newcommand{\code}[1]{\texttt{#1}}

\newcommand{\minPst}{minP}
\newcommand{\maxRew}{maxR}
\newcommand{\minRew}{minR}

\newcommand{\Value}[1]{\mathit{Val}\left[#1\right]}

\newcommand{\dMax}{\overline{\delays}}
\newcommand{\sFD}{S_{fd}}
\newcommand{\stepBound}{Bound[\#]}
\newcommand{\dOne}{D_1}

\newcommand{\maxValue}{\overline{\mathit{Val}}}

\begin{document}

\title{
Extension of PRISM by Synthesis of \\ Optimal Timeouts in Fixed-Delay CTMC
}

\titlerunning{Extension of PRISM by Synthesis of Optimal Timeouts in fdCTMC} 

\author{
\v{L}ubo\v{s} Koren\v{c}iak
\and Vojt\v{e}ch~{\v{R}}eh\'ak
\and Adrian Farmadin
}

\institute{
	Faculty of Informatics, Masaryk University, Brno, Czech Republic
		\texttt{\{korenciak,\,rehak,\,xfarmad\}\!@fi.muni.cz}
}

\maketitle 

\begin{abstract}
  We present a practically appealing extension of the probabilistic model
  checker PRISM rendering it to handle fixed-delay continuous-time Markov
  chains (fdCTMCs) with rewards, the equivalent formalism to the deterministic and stochastic
  Petri nets (DSPNs). fdCTMCs allow transitions with fixed-delays (or
  timeouts) on top of the traditional transitions with exponential
  rates. Our extension supports an evaluation of expected reward until reaching
  a given set of target states. The main contribution is that, considering the
  fixed-delays as parameters, we implemented a synthesis algorithm that
  computes the epsilon-optimal values of the fixed-delays minimizing the
  expected reward. We provide a performance evaluation of the synthesis on
  practical examples.

\end{abstract}

\section{Introduction}
\label{sec-intro}

PRISM~\cite{KNP:prismCAV11} is an efficient tool for probabilistic model-checking of stochastic
systems such as Markov decision processes (MDPs), discrete-time Markov chains (DTMCs), or continuous-time Markov
chains (CTMCs).  
The PRISM community frequently raises requests to incorporate the
possibility to express delays with deterministic durations in a
CTMC.\footnote{\url{http://www.prismmodelchecker.org/manual/FrequentlyAskedQuestions/PRISMModelling\#det\_delay}} The
standard PRISM recommendation is to approximate the deterministic durations
using a phase-type technique \cite{Neuts81} and thus obtaining a~CTMC. This
works for some models, however there are models for which such approximation
can cause either a large error or a state space explosion (see,
e.g. \cite{KKR:EPEW2014,fackrell2005fitting}). However, there is a formalism
called fixed-delay CTMCs (fdCTMCs) 
\cite{guet2012delayed,KKR:EPEW2014,BKKNR:QEST2015} that is the requested
extension of CTMCs by fixed-delay (fd) events, modeling the
deterministic transitions or timeouts. Recent result \cite{BKKNR:QEST2015}
came up with new synthesis algorithms working directly on fdCTMCs (rather
than approximating them with CTMCs). Here we provide the first attempt to
experimental evaluation of such synthesis algorithms and show that they are
practically applicable. In the following running example we demonstrate the
fdCTMC semantics as well as the parameters and objectives of the synthesis.

\begin{example}
  The figure bellow depicts fdCTMC of a slightly modified model of dynamic
  power management of a Fujitsu disk drive taken from the PRISM case
  studies\footnote{\url{http://www.prismmodelchecker.org/casestudies/power\_ctmc3.php}} \cite{QWP99}. The disk has three modes $\idle$, $\busy$,
  and $\sleep$. In the $\idle$ and $\sleep$ modes the disk receives requests, in the $\busy$
  mode it also serves them. The disk is equipped with a bounded buffer,
  where it stores requests when they arrive. The requests arrive with
  an exponential inter-arrival time of rate $1.39$ and increase the current
  size of the buffer. The requests are served in an exponential time of rate
  $12.5$, what decreases the buffer size. Note that restricting the model to
  the $\idle$ and $\busy$ modes only, we obtain a CTMC model of an M/M/1/n
  queue.

  Moreover, the disk can move from the $\idle$ mode to the $\sleep$ mode where it saves
  energy. 
  Switching of the disk to the
  $\sleep$ mode is driven by timeout. This is modeled by an \fd event $f_1$
  moving the state from $(\idle,0)$ to $(\sleep,0)$ when the disk is
  steadily idle for a specified amount of time (e.g. 1 second). The disk is
  woken up by another timeout modeled by an \fd event $f_2$, which is active
  in all $\sleep$ states. After staying in the $\sleep$ mode for, e.g. $2$
  seconds, $f_2$ changes the state according to the dashed~arrows.

\newcommand{\bend}{15}
  \begin{center}
  \begin{tikzpicture}[outer sep=0.1em, xscale=1, yscale=1]
      \tikzstyle{fixed}=[dashed,->]; \tikzstyle{fixed label}=[font=\small];
      \tikzstyle{exp}=[->,rounded corners,,>=stealth]; \tikzstyle{exp rate}=[font=\small];
      \tikzstyle{loc}=[draw,circle, minimum size=3.2em,inner sep=0.1em];
      \tikzstyle{accepting}+=[outer sep=0.1em]; \tikzstyle{loc
        cost}=[draw,rectangle,inner sep=0.07em,above=6, minimum
      width=0.8em,minimum height=0.8em,fill=white,font=\footnotesize];
      \tikzstyle{trans cost}=[draw,rectangle,minimum width=0.8em,minimum
      height=0.8em,solid,inner sep=0.07em,fill=white,font=\footnotesize];
      \tikzstyle{prob}=[inner sep=0.03em, auto,font=\footnotesize];

      \node[loc] (b0) at (0,0) {${\idle,0}$}; 
      \node[loc] (s0) at (0,-1.65) {${\sleep,0}$};

      \node[loc] (b1) at (2.5,0) {${\busy,1}$}; 
      \node[loc] (s1) at (2.5,-1.65) {${\sleep,1}$};

      \node[loc] (b2) at (5,0) {${\busy,2}$}; 
      \node[loc] (s2) at (5,-1.65) {${\sleep,2}$};

      \node[] (b3) at (7.5,0) {$\cdots$}; 
      \node[] (s3) at (7.5,-1.65) {$\cdots$};

      \node[loc] (bn) at (10,0) {$\busy, n$}; 
      \node[loc] (sn) at (10,-1.65) {$\sleep, n$};

      \path[->,>=stealth] ($(b0)+(-0.9,0)$) edge (b0);

\path[exp, bend left=\bend] (b0) edge node[prob] {1.39} 
(b1); 

\path[exp, bend left=\bend] (b1) edge node[prob] {12.5}
(b0);

\path[exp, bend left=\bend] (b1) edge node[prob] {1.39} 
(b2); 

\path[exp, bend left=\bend] (b2) edge node[prob] {12.5}
(b1);

\path[exp, bend left=\bend] (b2) edge node[prob] {1.39} 
(b3); 

\path[exp, bend left=\bend] (b3) edge node[prob] {12.5}
(b2);

\path[exp, bend left=\bend] (b3) edge node[prob] {1.39} 
(bn); 

\path[exp, bend left=\bend] (bn) edge node[prob] {12.5}
(b3);

\path[loop right,exp,looseness=5] (bn) edge node[prob, above, pos=0.2] {1.39}
(bn);

\path[exp] (s0) edge node[prob] {1.39} 
(s1); 

\path[exp] (s1) edge node[prob] {1.39} 
(s2); 

\path[exp] (s2) edge node[prob] {1.39} 
(s3); 

\path[exp] (s3) edge node[prob] {1.39} 
(sn); 

\path[loop right,exp,looseness=5] (sn) edge node[prob, above, pos=0.2] {1.39}
(sn);

\path[bend right=\bend,fixed] (b0) edge node[prob,left] {$f_1$}
(s0);

\path[bend right=\bend,fixed] (s0) edge node[prob, right] {$f_2$}
(b0);

\path[fixed] (s1) edge node[prob,right] {$f_2$}
(b1);

\path[fixed] (s2) edge node[prob,right] {$f_2$}
(b2);

\path[fixed] (sn) edge node[prob,right] {$f_2$}
(bn);

  \end{tikzpicture}
  \end{center}
  \label{fig:dpmsleep} 

  Additionally, every state is given a rate cost that specifies an amount of
  energy consumed per each second spent there. 
  Optionally, an impulse cost
  can be specified, e.g., say that the change from $(\idle,0)$ to $(\sleep,0)$
  consumes 0.006 energy units instantaneously. 
  Now, one might be interested in how much energy on average is consumed
  before emptying the buffer,  
  i.e. to compute the expected energy
  consumed until reaching target that is a new successor of $(\busy,1)$ instead of the initial state $(\idle,0)$. 
  But, being a developer of the disk, can we set better timeouts for $f_1$
  and $f_2$? Hence, we consider timeouts as parameters and synthesize them in order to minimize the expected amount of consumed energy.
\end{example}

\paragraphskip
\noindent \textit{Our Contribution} is as follows. 1.~We provide an
extension of the PRISM language and of the internal data structures to
support specification of fdCTMC with impulse and rate costs (or equivalently
rewards). Hence, our version of PRISM is now ready for other experiments
with fdCTMC algorithms including the possibility to support model-checking
options as for CTMCs and DTMCs. 2.~We added an evaluation of expected reward until reaching a given set of target states.
3.~We analyzed the synthesis algorithm from
\cite{BKKNR:QEST2015}, derived exact formulas and implemented the algorithm.
4.~Additionally, we accelerated the implementation by few structural changes, that significantly improved the running time and
the space requirements of the synthesis implementation. 5.~We provide a 
performance evaluation proving that
current implementation is practically applicable to a complex model from the PRISM case-study. 

\paragraphskip
\noindent \textit{Related Work { }} 
There are many papers that contain models with \fd events suitable for
synthesis such as deterministic durations in train control systems
\cite{Z:ECTS_synthesis}, time of server rejuvenation \cite{german-book},
timeouts in power management systems \cite{QWP99}, etc. Some of the models
already contain specified impulse or rate costs.

In \cite{XSCT:TRIVEDI_timeout_synthesis} authors compute the optimal value
of webserver timeout using impulse and rate costs. The implementation can
dynamically change the optimal value of timeout based on the current
inter-arrival times of requests. It works on the exact fdCTMC model and
cannot be easily applied to the more general fdCTMC models our implementation
can handle.

The formalism of deterministic and stochastic Petri nets (DSPNs) is  equivalent to fdCTMCs. DSPNs have been extensively studied and many useful
results are directly applicable to fdCTMCs. To the best of our knowledge the
synthesis of \fd events has not been studied for DSPNs.  The most useful tools for
DSPNs are ORIS \cite{HPRV:SSC} and TimeNET \cite{timenet}.

There was also an option to implement the synthesis algorithm as an
extension of ORIS. However, PRISM is much more used in practice and contains
solution methods for MDPs, that we needed for our implementation.
Thus, we decided to implement the synthesis into PRISM, even thought we had
to extend the PRISM language and data structures.
Therefore, the ORIS and TimeNET algorithms can be now reimplemented for
fdCTMCs in PRISM easily, exploiting its efficient symbolic structures and
algorithms for CTMCs or MDPs.

In the rest of the paper we first formally define the fdCTMC and explain the
extension of PRISM language. Then we discuss the implemented algorithms and
the performance results.

\section{Preliminaries}
\label{sec-prelims}

We use $\Nseto$, $\Rsetpo$, and $\Rsetp$ to denote the set of all
non-negative integers, non-negative real numbers, and positive real numbers,
respectively. Furthermore, for a countable set $A$, we denote by $\dist(A)$ the set of discrete probability distributions over $A$, i.e. functions $\mu: A \to \Rsetpo$ such that $\sum_{a\in A} \mu(a) = 1$.

\begin{definition}
  A \emph{fixed-delay CTMC} (fdCTMC) $\fdC$ is a tuple
  $(\states, \rates, \fDSet, \sched, \occur, \delays, \initstate)$ where
  \begin{itemize}
  \item $\states$ is a finite set of states,
  \item $\rates: \states \times \states \to \Rsetpo$ is a rate matrix,
  \item $\fDSet$ is a finite set of fixed-delay (fd) events,
  \item $\sched : \states \to 2^{\fDSet}$ assigns to each state $s$ a set of active \fd events in $s$,
  \item $\occur : \states \times \fDSet \to \dist(\states)$ is the successor function, i.e. assigns a probability distribution specifying the successor state to each state and \fd event that is active there,
  \item $\delays: \fDSet \to \Rsetp$ is a delay vector that assigns a
    positive delay to each \fd event,
  \item $\initstate \in \states$ is an initial state.
\end{itemize}
\end{definition}

Note that fdCTMC $\fdC$ with empty set of \fd events is a CTMC.
The fdCTMC formalism can be understood as a stochastic event-driven system,
i.e. the amount of time spent in each state and the probability of moving to
the next state is driven by the occurrence of events. In addition to the \fd
events of $\fDSet$, there is an \emph{exponential event} $\expEvent$ that is
active in all states $s$ where $\sum_{s' \in \states} \rates(s,s') >
0$. During an execution of an fdCTMC all active events keep one \emph{timer},
that holds the remaining time until the event occurs. The execution starts
in the state $\initstate$. The timer of each \fd event $f$ in
$\sched(\initstate)$ is set to $\delays(f)$. The timer of the exponential
event is set randomly according to the exponential distribution with a rate
$\sum_{s' \in \states} \rates(\initstate,s')$. The event $e$ with least\footnote{For the sake of simplicity, when multiple
  events $X = \{e_1,\ldots,e_n\}$ occur simultaneously, the successor is
  determined by the minimal element of $X$ according to some fixed total
  order on~$\fDSet$.} 
timer value $t$ occurs and causes change of state. In case $e$ is an \fd
event, the next state is chosen randomly according to the distribution
$\occur(\initstate,e)$, otherwise $e$ is an exponential event and the
probability of choosing $s$ as a next state is
$\rates(\initstate,s)/\sum_{s' \in \states}
\rates(\initstate,s')$. In the next state $s$, the timers of all newly active
\fd events (i.e. $\sched(s) \setminus \sched(\initstate)$), the occurred
event $e$, and the exponential event are set in the same way as above. Observe
that the timers of the remaining active \fd events decreased by time $t$
spent in the previous state. The execution then proceeds in the same manner.

We illustrate the definition on the fdCTMC model from
Example~\ref{fig:dpmsleep}. The execution starts in $(\idle,0)$. The events $f_1$ and $\expEvent$ are active and their timers are set to $1$ and e.g. $1.18$, respectively. Hence, after $1$ second $f_1$ occurs and changes the state to $(\sleep, 0)$
with probability $1$. The timers of newly active event $f_2$ and $\expEvent$  are set to $2$ and e.g. $1.5$, respectively. Now, $\expEvent$ occurs
and changes the state to $(\sleep,1)$. Here $f_2$ is still active and thus its timer holds the original value subtracted by the time spent in $(\sleep,0)$,
i.e. $2-1.5=0.5$. The timer of the exponential event is set, etc.

A \emph{run} of the fdCTMC 
is an infinite sequence $(s_0, e_0, t_0)(s_1,
e_1, t_1)\ldots$ where $s_0 = \initstate$ and for each $i \in \Nseto$ it
holds that $s_i \in \states$ is the $i$-th visited state, $e_i \in \{ \expEvent \} \cup \fDSet$ is the
event that occurred in $s_i$, and $t_i \in \mathbb{R}_{\geq 0}$ is the time
spent in $s_i$. For the formal definition of the semantics of fdCTMC
 and the probability space on runs see \cite{krcal_phd_thesis}. 

\paragraphskip
\noindent \textit{Total Reward Before Reaching a Target{ }}  To allow formalization of performance properties we enrich the model in a standard way with rewards or costs (see,~e.g.~\cite{Puterman:book}).
For an fdCTMC $\fdC$ with a state space $\states$ we additionally define a set
of target states $\goalStates$, reward rates $\rateRew$, and impulse rewards
$\impRew$.
Formally, the target state $\goalStates$ is a subset of $S \setminus
\initstate$, $\rateRew: \states \to \Rsetpo$ assigns a reward rate to every
state, and $\impRew: \states \times (\{ \expEvent \} \cup \fDSet) \times
\states \to \Rsetpo$ assigns an impulse reward
to every change of state.
Now the reward assigned to a run $(s_0, e_0, t_0)(s_1, e_1, t_1)\ldots$ is the
reward accumulated before reaching a state of $\goalStates$,
i.e. $\sum_{i=0}^{n-1} \left( t_i\cdot\rateRew(s_i) +
  \impRew(s_i,e_i,s_{i+1}) \right)$ where $n>0$ is the minimal index such
that $s_n\in\goalStates$. We set the reward to infinity whenever there is no
such~$n$.
The reward of a run can be viewed as a random variable, say $\costFdC$.
By $\expred_{\fdC,\goalStates,\rateRew,\impRew}$ (or simply $\expred_{\fdC}$) we denote the expected value of $\costFdC$.

\paragraph{Synthesis } Given a delay vector $\delays'$, let (parametric) fdCTMC
$\fdC(\delays')$ be the fdCTMC $\fdC$ where the delay vector is changed to
$\delays'$. Our aim is to find a delay vector $\delays$ such that the
expected reward $\expred_{\fdC(\delays)}$ is minimal. Formally, given an error bound
$\eps >0 $ the synthesis algorithm computes delay vector $\delays$, such
that $\expred_{\fdC(\delays)} \leq \Value{\fdC} + \eps$, where
$\Value{\fdC}$ denotes the optimal reward $\inf_{\delays'}
\expred_{\fdC(\delays')}$.

\section{PRISM Language and User Interface Extension}
Each fdCTMC model file must begin with the keyword \code{fdctmc}.  For the
purpose of our synthesis and expected reward implementation, the set of target states has to be
specified by label \code{"target"},~e.g.\vspace{\zmensovatkoVertMezer}
\begin{center}
  \code{label "target" = s=2;}
\end{center}
\vspace{\zmensovatkoVertMezer}
The exponential event (the matrix $\rates$) is specified the same way as in
CTMC models of PRISM. The \fd events are local to a module and must be
declared immediately after the module name. E.g. the \code{fdelay f = 1.0}
defines the \fd event $f$ with delay of a double value $1.0$. 
For an  \fd event $f$ we specify its set of active states (i.e. $\sched^{-1}(f)$)
and transition kernel  (i.e. $\occur(\cdot, f)$) by PRISM commands where the
identifier $f$ is in the arrow. E.g.
\vspace{\zmensovatkoVertMezer}
\begin{center}
  \code{[L] s=1 {-}{-}f-> 0.3:(s'=0) + 0.7:(s'=2)}
\end{center}
\vspace{\zmensovatkoVertMezer}
specifies that the \fd event $f$ is active in all states where \code{s=1}
and whenever it occurs, the next state is derived from the original one by
changing variable \code{s} to \code{0} with probability $0.3$ and to
\code{2} with probability $0.7$. The probabilities in each command have to
sum to one.  Observe that \fd event commands are similar to DTMC
commands in PRISM. The synchronization labels are used only to impose
impulse rewards as for CTMC, e.g.
\vspace{\zmensovatkoVertMezer}
\begin{center}
 \code{ rewards} ~~~ \code{[L] true : 1.0;}~~~\code{ endrewards}
\end{center}
\vspace{\zmensovatkoVertMezer}
 The rate rewards are
specified the same way as for CTMC in PRISM. 
The PRISM source code for
the fdCTMC of Example~\ref{fig:dpmsleep} is in
Appendix~\ref{app:prism-lang}. The implementation details concerning the fdCTMC
structure are provided in Appendix~\ref{app:prism-lang} as well.

Users can run the implemented algorithms from both the graphical and the command-line interfaces of PRISM. The expected reward and synthesis implementations are available in
menu \code{Model -> Compute -> Exp. reachability reward} and \code{Model -> Compute -> FD synthesis}, respectively or using the command-line option \code{-expreachreward} and 
\code{-fdsynthesis}, respectively.
The error bound $\eps$ is specified in
\code{Options -> Options -> Termination epsilon} or in the command-line option \code{-epsilon}.

\section{Implementation Issues}

Implementation of the expected reward computation was a straightforward application of existing PRISM methods. For the synthesis we implemented the \emph{unbounded optimization} algorithm from \cite{BKKNR:QEST2015}.
The algorithm is based on
discretization, i.e. we provide discretization bounds and restrict the uncountable space of
delay vectors into a
finite space. Instead of an exhaustive search through the finite space, we
use the idea of \cite{BKKNR:QEST2015} and transform the parametric
(discretized) fdCTMC into an MDP where actions correspond to the choices of
\fd event delays. Now, the minimal solution of the MDP yields the optimal delay vector.

The discretization bounds consist of the discretization step $ \delta $, the upper
bound on \fd event delay $ \dMax $ and the precision $ \kappa $ for computation of action parameters. 
They are computed for each \fd event separately from the error bound $\eps$, the number of states, the minimal transition probability, and other fdCTMC model attributes. For more detail see Appendix~\ref{app:synthesis}. 
Note that in every fdCTMC model, the delays for all \fd events have to be specified. 
Applying these delays, we compute
the corresponding expected reward $\maxValue$ which is used as an upper
bound for the optimal reward. 
Then $\maxValue$ is employed when computing the
discretization bounds. 
The lower the $\maxValue$ is, the faster the
synthesis implementation performs. 
Thus it is worth to think of good delays
of \fd events when specifying the model.

Given the discretization bounds one has to compute the transition probabilities and expected accumulated reward for each action in the MDP corresponding to the discretized delay of \fd event. This can be done using the transient analysis of subordinated CTMCs~\cite{DBLP:journals/pe/Lindemann93}.

\paragraphskip
\noindent \textit{Prototype Implementation { } }In the first implementation
we used straightforward approach to call built-in methods of PRISM to compute
the required quantities for each discretized \fd event delay separately.
This is reasonable since the built-in methods are correctly and efficiently programmed for all PRISM engines and methods of computing transient analysis. 
However, we experienced that most of the time was spent computing the transient analysis rather than solving the created MDP, e.g. $ 520 $ seconds out of $ 540 $ seconds of total time.\footnote{Computed for the rejuv model and the error bound $ 0.001 $, see Section~\ref{sec:experiments}.} 
One of the reasons is that in each iteration a small portion of memory is allocated and freed by built-in PRISM methods. 
Since there is a large number of actions, the amount of reallocated memory was slowing down the computation.
Thus we decided to reimplement the computation of transient probabilities the applying principles of dynamic programming.

\paragraphskip
\noindent \textit{Iterative Computation of Transient Analysis { } }
The transient probabilities can be very efficiently approximated up to an arbitrary small error using the uniformization technique. 
The problem is that we have to compute the transient probabilities for each value of a very large set $\{i \cdot \delta \mid i \in \Nseto \text{ and } 0 < i \leq \dMax /\delta \}$ and allow only fixed error $\kappa$ for each computation. 
The transient probability vector $\pi(\delta)$ of a CTMC $\fdC$ at time $\delta$ can be computed using uniformization by 
\begin{equation}
\pi(\delta) = \sum_{j=0}^{J} \mathbf{1}_{\initstate} \cdot P^j \cdot \frac{(\lambda \cdot \delta)^j}{j!} \cdot e^{-\lambda \cdot \delta},
\end{equation}
where $\mathbf{1}_{\initstate}$ is the initial vector of $\fdC$, $\lambda$ is the uniformization rate of $\fdC$, and $P$ is the transition kernel of the  uniformized $\fdC$. 
The choice of number $J$ influences the error of the formula. 
It is easy to compute the value of $J$ such that the error is sufficiently small.

However, for time $i\cdot \delta $ we can use the previously computed transient probabilities as 
\begin{equation}\label{eq:iterative1}
\pi(i \cdot \delta) = \sum_{j=0}^{J} \pi((i-1) \cdot \delta) \cdot P^j \cdot \frac{(\lambda \cdot \delta)^j}{j!} \cdot e^{-\lambda \cdot \delta}.
\end{equation}
It is again easy to compute $J$ such that the overall allowed error is not exceeded. 
Instead of performing na\"{i}ve computation for each number in $\{i \cdot \delta \mid i \in \Nseto \text{ and } 0 < i \leq \dMax /\delta \}$
with according number of steps $J_1, \ldots, J_{\dMax/\delta}$ to cause error bounded by $\kappa$
in each computation, we compute the transient probabilities
iteratively with sufficiently large $J$ to cause small error in all
computations. For example, if we have $\delta = 0.1$, $\dMax/\delta=1000$, rate
$\lambda= 1.0$ and $\kappa = 0.01$ using the na\"{i}ve method we have to do $ J_1 + \cdots + J_{\dMax/\delta} = 66,265 $
steps and using the iterative method $J \cdot \dMax /\delta = 3,000$
steps. This is significant difference since a vector matrix multiplication is performed in each step. 
Thus we hard-programmed the iterative computation of transient probabilities and accumulated rewards in CTMC what caused a dramatic speedup thanks to the smaller number of arithmetic operations and better memory management.

\paragraphskip
\noindent \textit{Precomputation{ } }Careful reader may have noticed that \eqref{eq:iterative1} can be further simplified to 
\begin{equation} \label{eq:iterative2}
\pi(i \cdot \delta) = \pi((i-1) \cdot \delta) \cdot e^{-\lambda \cdot \delta} \cdot \sum_{j=0}^{J} P^j \cdot \frac{(\lambda \cdot \delta)^j}{j!}.
\end{equation}
Hence, the matrix $ e^{-\lambda \cdot \delta} \cdot \sum_{j=0}^{J} P^j \cdot {(\lambda \cdot \delta)^j}/{j!} $ can be easily precomputed beforehand and used for computation of each $ \pi(i \cdot \delta) $ to increase the savings even more. However, this is not true. 
$ J $ is small and the matrix $ P $ is sparse for the most reasonable models and error bounds. 
But $ e^{-\lambda \cdot \delta} \cdot \sum_{j=0}^{J} P^j \cdot {(\lambda \cdot \delta)^j}/{j!} $ is not sparse for almost each error bound, $ P $, and $ \lambda $, what is known as "fill-in" phenomenon. 
Thus using \eqref{eq:iterative1} is typically more efficient than using \eqref{eq:iterative2}.
Similar observations were discussed in \cite{HMM:transien_analysis_DSPN}.

Implementing the synthesis algorithm of \cite{BKKNR:QEST2015}, we inherited
the following restrictions on the input fdCTMC models.
There is at most one concurrently active \fd event in
each state, i.e. $\forall s \in \states \, : \, |\sched(s)| \leq 1$.
For each \fd event there is at most one state
where its timer is set.
Every state has a positive rate reward, i.e. $\forall s \in \states \, : \,
\rateRew(s) > 0$. 
Moreover,  we add that all \fd events have positive impulse rewards,
i.e. $ \forall f \in \fDSet \wedge s,s' \in \states : \occur(s,f)(s') >0 \implies \impRew(s,f,s') > 0$.  
For the expected reward implementation only the first two restrictions are valid.

\section{Experimental Results}\label{sec:experiments}
We tested the performance of our synthesis implementation on the model from
Example~\ref{fig:dpmsleep} for various sizes of the queue ($2,4,6$, and $8$)
and the rejuvenation model provided in Appendix~\ref{app:prism-lang}.  The
considered error bounds are $0.005$, $0.0025$, $0.0016$, $0.00125$, and
$0.001$. The following table shows the expected rewards and the computation
times for a given error bound. As the expected rewards are very similar for
different error bounds, we show their longest common prefix, instead of
listing five similar long numbers.
\begin{center}
\begin{tabular}{| c | l  r | r | r | r | r | l |}
\hline
 \multirow{3}{*}{ Model } 
& \multicolumn{6}{c|}{CPU time [s]}  
& \multicolumn{1}{c|}{Longest} \\
\cline{2-7}
&~ $\eps:$  ~& 0.005 ~&~ 0.0025 ~&~ 0.0016 ~&~ 0.00125 ~&~ 0.00100 ~& \multicolumn{1}{c|}{ common prefix }
\\
\cline{2-7} 
&~ $ 1/\eps:$ ~& 200  ~&~ 400  ~&~ 600  ~&~ 800  ~&~ 1000 ~&  \multicolumn{1}{c|}{of exp. rewards}
\\ \hline
~rejuv ~& \multicolumn{2}{r|}{ 5.87 ~} &~ 12.09  ~&~ 14.71  ~&~ 21.60  ~&~ 23.84  ~&~ 0.94431314832 ~\\
~dpm2 ~& \multicolumn{2}{r|}{ 58.22 ~} &~ 121.15  ~&~ 195.61  ~&~ 234.58  ~&~ 248.52  ~&~ 0.336634754 ~\\
~dpm4 ~& \multicolumn{2}{r|}{ 156.02 ~} &~ 354.35  ~&~ 509.19  ~&~ 2197.10  ~&~ 2652.05  ~&~ 0.337592724 ~\\
~dpm6 ~& \multicolumn{2}{r|}{ 259.76 ~} &~ 532.47  ~&~ 2705.45  ~&~ 3026.77  ~&~ 5124.10  ~&~ 0.337583980 ~\\
~dpm8 ~& \multicolumn{2}{r|}{ 616.47 ~} &~ 3142.44  ~&~ 6362.79  ~&~ 22507.55  ~&~ 27406.62  ~&~ 0.337537611 ~\\
\hline
\end{tabular}

\end{center}
Note that the computed values of the expected reward are of a much better
precision than required. This indicates that there might even be a space for
improvements of the synthesis algorithm, e.g. by computation of tighter
discretization bounds.  It is worth mentioning that the longest computation
(dpm8 for error $0.001$) took only 1~hour and 30 minutes of real clock time
thanks to the native parallelism of Java (the table shows the
sum 
for all threads).  Our experiments show that the implementation retains the
theoretical complexity bounds saying that the computation time is
exponential to the number of states and polynomial to $1/\eps$.

The computations were run on platform HP DL980 G7 with 8 64-bit processors
Intel Xeon X7560 2.26GHz (together 64 cores) and 448 GiB DDR3 RAM, but only
304GB was provided to Java. The time was measured by the Linux command
\code{time}.

\section{Conclusions and Future Work
}\label{sec:concl}

In this paper, we incorporated the fdCTMC models into PRISM and implemented the expected reward computation and the synthesis algorithm.
The tool is
available on \url{http://www.fi.muni.cz/~xrehak/fdPRISM/}. 
We have used the explicit state PRISM engine. 
Based
on the promising results, it is reasonable to (re)implement the synthesis
and other model checking algorithms for fdCTMCs in the more efficient PRISM
engines. 
Moreover, new effort can be put to reduce the number of current restrictions
on the fdCTMC models. 
For instance the method of stochastic state classes~\cite{HPRV:SSC} implemented in ORIS may be applied for computation of transient analysis instead of uniformization.

\subsubsection{Acknowledgments} We thank Vojt\v{e}ch Forejt and
David Parker for fruitful discussions. This work is partly supported by the Czech Science Foundation, grant No.~P202/12/G061.

\vspace{-1.0\baselineskip}
\bibliographystyle{splncs03}
\bibliography{str-short,concur}
\newpage
\appendix
\section{Discretization Bounds}\label{app:synthesis}
Using the full version of \cite{BKKNR:QEST2015} we derived the exact
formulas of the discretization bounds for each \fd event:
\begin{align*}
&\dMax = \max\Big\{ \frac{\maxValue}{\minPst^{|\sFD|} \cdot \minRew} 
\; ; \; \frac{e \cdot | \ln(\alpha/2) | } {\lambda \cdot \minPst} 
\Big\}, \\ 
&\delta = \frac{\alpha}{\dOne}, \\
&\kappa = \frac{\eps \cdot \delta \cdot \minRew}{2 \cdot |S'| \cdot (1 + \maxValue)},
\end{align*}
where
\begin{align*}
&\alpha = \min\Big\{ \frac{\eps}{\stepBound \cdot (1 + \maxValue) \cdot |S'|} \; ; \; \frac{1}{2 \cdot \stepBound \cdot |S'|} \Big\}, \\
&\dOne = \max \{2 \cdot \lambda \; ; \; 1 \cdot (\lambda+1)\cdot \maxRew \}, 
\end{align*}
\begin{itemize}
	\item $ \stepBound $ is an upper bound on expected number of steps
          to reach target from any state in the created MDP, i.e.
$$\stepBound = \frac{\maxValue}{\text{minimal expected one-step reward in the created MDP}},$$
	\item $ \maxValue $ is the upper bound on the expected reward,
	\item $ S' $ is the state space of the created MDP, 
	\item $ \lambda $ is the uniformization rate, and
	\item $ |\sFD| $, $ \minPst $, $ \maxRew $, and $ \minRew $  is the
          number of states, the minimal branching probability, the maximal
          reward, and the minimal reward in the subordinated CTMC for the given
          \fd event, respectively. 
\end{itemize}

\section{fdCTMC in PRISM Language}\label{app:prism-lang}

\paragraphskip
\noindent \textit{Extension of PRISM data structures { }} The \code{FDCTMCSimple} class extends the \code{CTMCSimple} class by a vector of objects of type \code{FDEvent} and few methods to work with them (the methods are explained in the corresponding interface \code{FDCTMC} that is an extension of interface \code{CTMC}). The 
\code{FDEvent} is basically an extension of \code{DTMCSimple} class by one \code{double} attribute that keeps delay of the \fd event and one \code{String} attribute that keeps the label of the \fd event. The transition kernel is kept in the inherited attributes from the \code{DTMCSimple} class.

\begin{figure}
	\begin{center}
		\includegraphics[width=340pt]{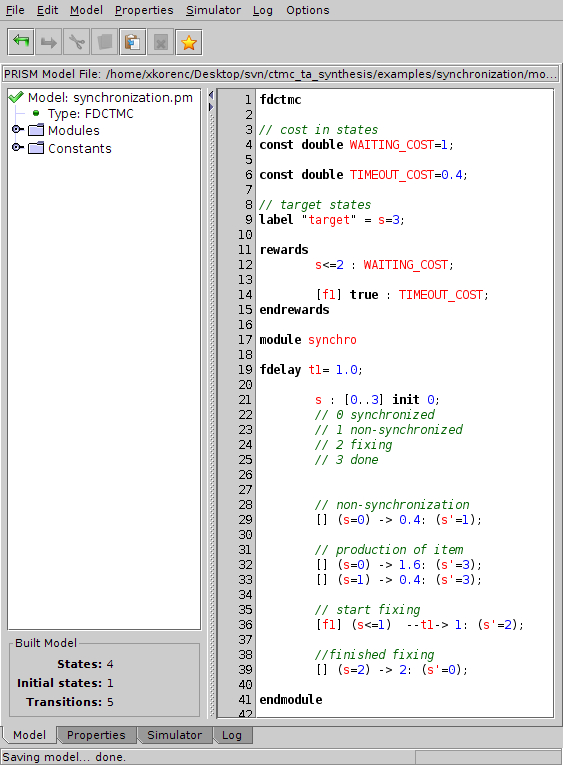}
		\caption{The graphical user interface of PRISM with the source code of the rejuvenation model \cite{german-book}.}
		\label{fig:rejuv}
	\end{center}
\end{figure}

\begin{figure} 
	\begin{center}
		\includegraphics[width=330pt]{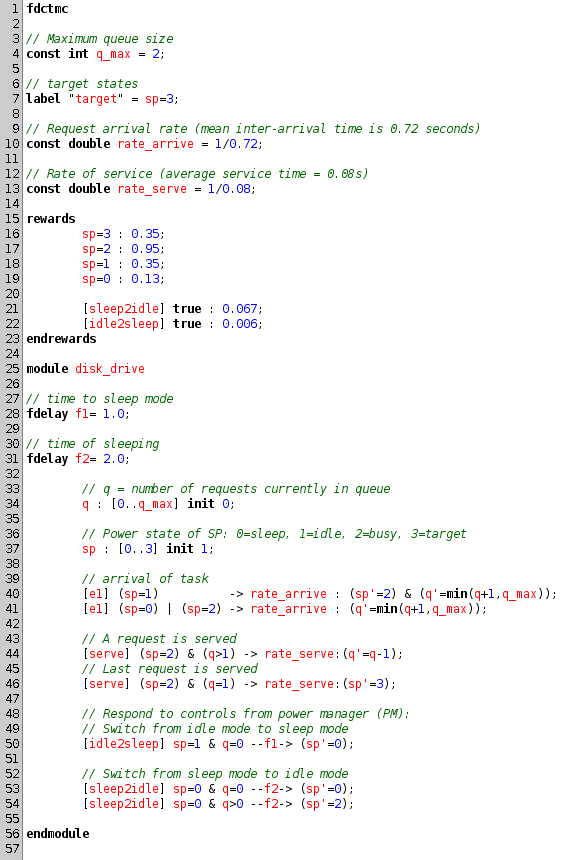}
		\caption{The source code of the fdCTMC from Example~\ref{fig:dpmsleep} in the PRISM language.}
		\label{fig:prism-lang}
	\end{center}
\end{figure}

\end{document}